\documentclass[twocolumn,pra,aps,showpacs,superscriptaddress,floatfix]{revtex4-1}
\usepackage[latin1]{inputenc}
\usepackage{bm}
\usepackage[usenames]{color}
\usepackage{multirow}
\usepackage{amssymb}
\usepackage{amsbsy}
\usepackage{amsmath}
\usepackage{stmaryrd}
\usepackage{graphicx}
\usepackage{epsfig}
\usepackage{placeins}
\usepackage{ulem}
\makeatletter

\newcommand{\tn}{\textnormal}

\begin{document}

\title{Spin and thermal conductivity of quantum spin chains and ladders}

\author{C. Karrasch} 
\affiliation{Department of Physics, University of California, Berkeley, California 95720, USA}
\affiliation{Materials Sciences Division, Lawrence Berkeley National Laboratory, Berkeley, CA 94720, USA}

\author{D. M. Kennes}
\affiliation{Institut f\"ur Theorie der Statistischen Physik, RWTH Aachen University and JARA-Fundamentals of Future Information Technology, 52056 Aachen, Germany}

\author{F. Heidrich-Meisner}
\affiliation{Department of Physics and Arnold Sommerfeld Center for Theoretical Physics,
Ludwig-Maximilians-Universit\"at M\"unchen, 80333 M\"unchen, Germany}

\date{\today}

\begin{abstract}
We study the spin and thermal conductivity of spin-1/2 ladders at finite temperature. This is relevant for 
experiments with quantum magnets. Using a state-of-the-art density matrix renormalization
group algorithm, we compute the current autocorrelation functions on the real-time axis and then carry out a Fourier integral to extract the frequency dependence of the corresponding conductivities. The finite-time error is analyzed carefully.
We first investigate the limiting case of spin-1/2 XXZ chains, for which our analysis suggests non-zero dc-conductivities in all interacting cases irrespective of the presence or absence of spin Drude weights. 
For ladders, we observe that all models studied 
are normal conductors with no ballistic contribution. 
Nonetheless, only the high-temperature spin conductivity of XX ladders has a simple diffusive, Drude-like form, while
 Heisenberg ladders exhibit a more complicated low-frequency behavior.
We compute the dc spin conductivity down to temperatures of the order of $T\sim 0.5J$, where $J$ is the exchange coupling along the legs of the ladder. We further extract 
mean-free paths and discuss  our results   in relation to thermal conductivity measurements on quantum
magnets.
\end{abstract}

\maketitle

\section{Introduction}

Low-dimensional quantum magnetism is a field in which an extraordinary degree of 
quantitative agreement between experimental results and theory has been achieved due to the availability of both high-quality samples and  powerful theoretical tools such as bosonization \cite{giamarchi}, Bethe ansatz \cite{kluemper-book},
series expansion methods \cite{knetter00,trebst00}, or the density matrix renormalization 
group \cite{white92,schollwoeck05}. This includes the thermodynamics 
\cite{johnston00a,johnston00}, inelastic neutron scattering data \cite{notbohm07,lake13},
as well as various other spectroscopic methods \cite{windt01}.
While there are also exciting experimental results for spin diffusion probed via NMR \cite{thurber01,branzoli11}   
or $\mu$sr \cite{maeter,pratt06} as well as for
the thermal conductivity \cite{sologubenko07,hess07},  the calculation of finite-temperature linear-response transport coefficients poses a formidable problem for theorists (see Refs.~\onlinecite{zotos-review,hm07} for a review), which is further complicated by the need to account for phonons and impurities
(see Refs.~\onlinecite{shimshoni03,chernyshev05,rozhkov05a,boulat07,gangadharaiah10,bartsch13,rezania13} for work in this direction).

Very recently, significant progress has been made in the computation  of linear response transport properties of the seemingly simplest one-dimensional model, the integrable spin-1/2 XXZ chain 
with an exchange anisotropy $\Delta$. Its Hamiltonian reads:
\begin{equation}\label{eq:hxxz}
H = J \sum_{n=1}^{L-1} \left[ S^x_{n}S^x_{n+1} + S^y_{n}S^y_{n+1} + \Delta S^z_{n}S^z_{n+1}\right]\,,
\end{equation}
where $S^{x,y,z}_{n}$ is a spin-1/2 operator acting on site $n$. The spin and thermal conductivities generally take the form
\begin{equation}\label{eq:sigma}\begin{split}
\mbox{Re}\,\sigma(\omega) &= 2\pi D_\tn{s\phantom{h}}\delta(\omega) + \sigma_\tn{reg}(\omega)\,,\\
\mbox{Re}\,\kappa(\omega) &= 2\pi D_\tn{th}\delta(\omega) + \kappa_\tn{reg}(\omega)\,,
\end{split}\end{equation}
where $D_\tn{s,th}$ denote the Drude weights, and $\sigma_\tn{reg}$ and $\kappa_\tn{reg}$ are the regular parts. The exact conservation of the energy current \cite{zotos97} of the XXZ chain renders 
the zero-frequency thermal conductivity  strictly divergent at all temperatures, i.e., $D_\tn{th}>0$, $\kappa_\tn{reg}=0$. The thermal Drude weight has been calculated exactly \cite{kluemper02,sakai03}. For spin transport, the following picture emerges: while there is a regular contribution $\sigma_\tn{reg}>0$ for all $|\Delta|>0$ \cite{naef98}, the Drude weight $D_\tn{s}$ is non-zero for $|\Delta|<1$ but vanishes for $|\Delta|>1$. Initially, these results were largely based on numerical simulations \cite{zotos96,narozhny98,hm03,heidarian07,herbrych11,karrasch12,karrasch13} as well
as analytical approaches using the Bethe ansatz \cite{zotos99,peres99,benz05}. 
Recently, a rigorous proof of finite spin Drude weights for $|\Delta|<1$ has been  obtained \cite{prosen11,prosen13} by relating $D_\tn{s}>0$ to the existence of a novel family of quasi-local conservation laws via the Mazur inequality \cite{zotos97}. For the experimentally most relevant case of the spin-1/2 Heisenberg chain ($\Delta=1$), it is still debated whether or not a ballistic contribution exists at finite temperatures (see Refs.~\onlinecite{herbrych11,karrasch13,steinigeweg14,carmelo14} for recent work).
The same questions of diffusive versus ballistic transport can be addressed in non-equilibrium setups \cite{langer09,jesenko11,langer11,karrasch14} or for open quantum systems \cite{prosen09,znidaric11,znidaric13a,mendoza-arenas13}.
A recent quantum gas experiment, in which the ferromagnetic Heisenberg chain was realized 
with a two-component Bose gas, studied the decay of a spin spiral, and the results were interpreted in
terms of diffusion \cite{hild14}. Other non-equilibrium experiments with quantum gases have investigated the 
mass transport of fermions \cite{schneider12} and bosons \cite{ronzheimer13} in optical lattices. 

Another very interesting question pertains to the functional form of the regular part $ \sigma_{\rm reg }(\omega)$. A field-theoretical study \cite{sirker09,sirker11}, which incorporates the leading irrelevant umklapp term, suggests that $ \sigma_{\rm reg }(\omega)$  has a simple diffusive form at low temperatures $T\ll J$. This is consistent with early results \cite{giamarchi91} for the generic behavior of a Luttinger liquid in the presence of umklapp scattering as well as with Quantum Monte Carlo simulations for $\Delta=1 $ \cite{grossjohann10}. At higher temperatures, a suppression of weight at low frequencies according to $ \sigma_{\rm reg }(\omega) \propto \omega^2$ has been  suggested \cite{herbrych12}. Most studies of $|\Delta|>1$ are interpreted in terms of diffusive spin dynamics, i.e., finite dc-conductivities 
 \cite{prosen09,steinigeweg09,znidaric11,steinigeweg12,karrasch14, karrasch14a}; however, indications of an anomalous low-frequency response were reported in Ref.~\onlinecite{prelovsek04}.
The theory by Sachdev and Damle provides a semi-classical interpretation for the emergence of diffusive dynamics in gapped spin models and predictions for the low-temperature dependence of the diffusion constant \cite{sachdev97,damle98,damle05}.

Many transport experiments on quantum magnets probe materials which are described by quasi-one dimensional models  more complicated than the integrable XXZ chain. Most notably, very large thermal conductivities due to spin excitations have been  observed in spin-ladder compounds \cite{sologubenko00,hess01, hess07}, which more recently have  also been investigated using real-time techniques \cite{otter09,montagnese13,hohensee14}. Most theoretical studies of non-integrable models  suggest the absence of ballistic contributions \cite{zotos96,rosch00,hm03,hm04,zotos04,karadamoglou04,jung06} (possible exceptions have been proposed in  Refs.~\onlinecite{karrasch12a,znidaric13,znidaric13a}). Numerical results for the expansion of local spin and energy excitations in real space are consistent with diffusive dynamics \cite{langer09,karrasch14}. A qualitatively similar picture has emerged from studies of transport in open quantum systems \cite{prosen09,mendoza-arenas14}. Despite the relevance for experiments, however, transport properties of generic non-integrable systems are still not fully understood quantitatively. Two important and largely open problems in the realm of spin ladders are (a) the question of whether they exhibit standard diffusive dynamics or a more complicated low-frequency behavior, and (b) a quantitative calculation of their dc spin and thermal conductivities. It turns out that a  Drude-like $\sigma_{\rm reg}(\omega)$ rarely exists in quasi one-dimensional spin Hamiltonians with short-range interactions (see, e.g.,  Refs.~\onlinecite{zotos04,hm04c}).  A notable example in which  standard diffusion is realized in the high-temperature regime is the XX spin ladder \cite{steinigeweg14a}, which is equivalent to hard-core bosons and thus relevant for recent experiments on mass transport of strongly interacting bosons in optical lattices in one and two dimensions \cite{ronzheimer13,vidmar13}. 

The main goal of our work is to compute the frequency dependence of the spin and thermal conductivity of spin ladders as well as of the spin-1/2 XXZ chain. We use a finite-temperature, real-time version of the density matrix renormalization group method (DMRG) \cite{karrasch12,barthel13,karrasch13a,kennes14} based on the purification trick \cite{verstraete04}.
 This method allows one to calculate both thermodynamics \cite{feiguin05} but also the time dependence of 
current autocorrelation functions. We calculate the conductivities from  Kubo formulae. 
For the accessible time scales, our results are free of finite-size effects \cite{karrasch13} and thus effectively describe systems in the thermodynamic limit.
Exploiting several recent methodological advances and using an optimized and parallelized implementation allows us to access larger time scales than in earlier applications of the method \cite{karrasch12,karrasch13,karrasch14}. Our data agree well with exact diagonalization approaches \cite{zotos04} for the thermal conductivity of spin ladders and the spin transport in XX ladders \cite{steinigeweg14a}. The latter results have been obtained from a pure state propagation method based on the dynamical typicality approach, which has recently been  applied to the calculation off 
transport coefficients  \cite{steinigeweg14,steinigeweg14a,steinigeweg14b}.

Our key results are as follows. For the spin-1/2 XXZ chain with $ 0<\Delta < 1$, we provide evidence that $\sigma_{\rm reg }(\omega)$ remains finite in the dc limit, but its low-frequency behavior is not of a simple Lorentzian form. For $\Delta=0.5$, we observe a suppression of weight for $\omega\ll J$ in the high-temperature regime.
In the case of spin ladders, $\sigma_{\rm reg}(\omega)$ also generically exhibits a complicated low-frequency dependence, and a simple Drude-like form is recovered only in the XX case $\Delta=0$ in agreement with the results of Ref.~\onlinecite{steinigeweg14a}. We extract the dc spin conductivity for temperatures $T \geq 0.5J$ and discuss how it depends on the exchange anisotropy $\Delta$. We translate the high-$T$ spin and thermal conductivities of the Heisenberg ladder into mean-free paths by fitting to a simple phenomenological expression often used in the interpretation of  experimental data \cite{hess01}. It turns out that the values of the mean-free paths depend on which type of transport is considered.

The structure of this exposition is as follows. We introduce the model and definitions in Sec.~\ref{sec:def}.
Section~\ref{sec:num}  provides details on our numerical method.
Our results are summarized in Sec.~\ref{sec:results}, where we discuss the real-time dependence of 
current correlations and the methods to convert them into frequency-dependent conductivities, which we then study for spin chains and ladders.
Our conclusions are presented in Sec.~\ref{sec:sum}.

\section{Model and definitions}
\label{sec:def}
The prime interest of this work is in two-leg spin ladders governed by the Hamiltonian $H=\sum_{n=1}^{L-1} h_n$ and local terms 
\begin{equation}\label{eq:hlad}\begin{split}
h_n = J & \sum_{\lambda=1,2} \left[ S^x_{n,\lambda}S^x_{n+1,\lambda} + S^y_{n,\lambda}S^y_{n+1,\lambda} + \Delta S^z_{n,\lambda}S^z_{n+1,\lambda}\right] \\
+ \frac{J_\perp}{2} & \sum_{m=n,n+1} \left[S^x_{m,1}S^x_{m,2} + S^y_{m,1}S^y_{m,2} + \Delta S^z_{m,1}S^z_{m,2}\right]\,,
\end{split}\end{equation}
where $S^{x,y,z}_{n,\lambda}$ are spin-$1/2$ operators acting on the rung $\lambda=1,2$. The model is non-integrable and gapped for all $J_\perp>0$. At $J_\perp=0$, one recovers two identical, decoupled XXZ chains, which (at zero magnetization) are gapless for $|\Delta|\leq1$ and gapped otherwise \cite{giamarchi}.

Both the Drude weights and the regular parts of the spin (s) and thermal (th) conductivities defined in Eq.~(\ref{eq:sigma}) can be obtained from the corresponding current correlation function $C_\tn{s,th}(t)$. Their long-term asymptote is related to $D_\tn{s,th}$ via
\begin{equation}\begin{split}\label{eq:dw}
D_\tn{s,th} & = \lim_{t\to\infty}\lim_{L\to\infty} \frac {C_{\rm s,th}(t) }{2T^{\alpha_\tn{s,th}}},~
C_{\rm s,th}(t) = \frac{\mbox{Re} \, \langle I_\tn{s,th}(t) I_\tn{s,th}\rangle}{L},
\end{split}\end{equation}
where $\alpha_\tn{s}=1$ and $\alpha_\tn{th}=2$. The regular part of the conductivity is determined by
\begin{equation}\label{eq:sigma1}\begin{split}
&\tn{Re}\,\left\{{{\sigma_\tn{reg}(\omega)}\atop{\kappa_\tn{reg}(\omega)}}\right\} =\, \frac{1-e^{-\omega/T}}{\omega T^{\alpha_\tn{s,th}-1} }\times \\
&~~~~~\tn{Re} \int_0^{\infty}dte^{i\omega t} \lim_{L\to\infty}\left[ C_\tn{s,th}(t) - 2T^{\alpha_\tn{s,th}}D_\tn{s,th}\right]\,.
\end{split}\end{equation}
Only finite times can be reached in the DMRG calculation of $C_\tn{s,th}(t)$, which leads to a `finite-time' error of $\sigma_\tn{reg}(\omega)$ that can be assessed following Ref.~\onlinecite{karrasch14a}. We will elaborate on this below.

The current operators $I_\tn{s,th} = \sum_n j_{(\tn{s,th}),n}$ are defined via the respective continuity equations \cite{zotos97}. 
The local spin-current operators of the XXZ chain take the well-known form $j_{\textnormal{s},n} = i J S^x_{n} S^y_{n+1} +\tn{h.c.}$. For the spin ladder, one finds
\begin{eqnarray}
 j_{\textnormal{s},n} &=& i J \sum_\lambda \big(S^x_{n,\lambda} S^y_{n+1,\lambda} - S^y_{n,\lambda}S^x_{n+1,\lambda} \big)\,, \\
j_{\textnormal{th},n} &=& i [h_{n}, h_{n+1}]\,.
\end{eqnarray}
Note that our definition for the local energy density $h_n$ preserves all spatial symmetries of the ladder, and our energy-current operator $I_\tn{th}$ is the same as the one used in Ref.~\onlinecite{zotos04}. The full expression for $I_{\textnormal{th}}$ is lengthy and not given here.

\section{Numerical method}
\label{sec:num}

We compute the spin- and energy-current correlation function
\begin{eqnarray}\label{eq:time}
 \langle I_\tn{s,th}(t) I\rangle &\sim& \tn{Tr}\big[ e^{-H/T}e^{iHt}I_\tn{s,th}e^{-iHt}I_\tn{s,th}\big]
\end{eqnarray}
using the time-dependent \cite{vidal04,white04,daley04,schmitteckert04,vidal07} density matrix renormalization group \cite{white92,schollwoeck05} in a matrix-product state \cite{fannes91,ostlund91,verstraete06,verstraete08,schollwoeck11} implementation. Finite temperatures \cite{verstraete04,white09,barthel09,zwolak04,sirker05,white09,barthel13} are incorporated via purification of the thermal density matrix. 
Purification is a concept from quantum information theory in which the physical system is embedded into an environment. The wave-function of
the full system is then a pure state and the mixed state describing the system is obtained by tracing out the degrees of freedom
of the environment. 
When using this approach in  DMRG, one typically simply chooses a copy of the system degrees of freedom to be the environment.
Details of purification based finite-$T$ DMRG methods can be found in Refs.~\cite{verstraete04,feiguin05,schollwoeck11,barthel13,karrasch14}. Our actual implementation follows
Ref.~\cite{karrasch14}.

The real- and imaginary time evolution operators in Eq.~(\ref{eq:time}) are factorized by a fourth-order Trotter-Suzuki decomposition with a step size of $dt=0.05,\ldots, 0.2$. We keep the discarded weight during each individual `bond update' below a threshold value $\epsilon$. This leads to an exponential increase of the bond dimension $\chi$ during the real-time evolution. In order to access time scales as large as possible, we employ the finite-temperature disentangler introduced in Ref.~\onlinecite{karrasch12}, which uses the fact that purification is not unique to slow down the growth of $\chi$. Moreover, we `exploit time translation invariance' \cite{barthel13},  rewrite $\langle I_\tn{s,th}(t)I_\tn{s,th}(0)\rangle=\langle I_\tn{s,th}(t/2)I_\tn{s,th}(-t/2)\rangle$, and carry out two independent calculations for $I_\tn{s,th}(t/2)$ as well as $I_\tn{s,th}(-t/2)$. Our calculations are performed 
using a system size of $L=100$ for spin ladders and $L=200$ for the XXZ chain, respectively. By comparing to  other values of $L$ we have ensured that $L$ 
is large enough for the results to be effectively in the thermodynamic limit \cite{karrasch13}.

\section{Results}
\label{sec:results}

\subsection{Current autocorrelation functions}
\label{sec:current}
 
Figure~\ref{Fig1} shows typical results for the decay of  spin-current autocorrelations of the XXZ chain 
as a function of time. Some of these data have previously been shown in Refs.~\onlinecite{karrasch13,karrasch14,karrasch14a} and are here included for comparison. For $\Delta<1$, we clearly observe the 
saturation of $C_{\rm s}(t)$ at a non-zero value at long times that at $T=\infty$ agrees well with an  improved lower bound \cite{prosen13} for $\lim_{T\to\infty} TD_\tn{s}(T)$ 
and Zotos' Bethe-ansatz calculation \cite{zotos99}. 
 At $\Delta=0.5$, the values for $D_\tn{s}$ obtained in Ref.~\onlinecite{zotos99} coincide with our tDMRG data also for the finite temperatures $T< \infty$ considered here 
(see the inset to Fig.~\ref{Fig1}; compare Ref.~\onlinecite{karrasch13}).
In the case of $\Delta>1$, the current correlators appear to decay to zero, consistent with predictions of a vanishing finite-temperature Drude weight in this regime \cite{zotos96,peres99,hm03,karrasch14}.
At the isotropic point $\Delta=1$, $C_{\rm s}(t)$ does not saturate to a constant on the time scale reached in the simulations \cite{karrasch13}, and no conclusion on the presence or absence of a ballistic contribution is possible.

We next turn to the case of spin ladders. Exemplary DMRG data for $C_{\rm s}$ and $C_{\rm th}$ at three different temperatures $T\in\{\infty,J,0.5J\}$ are shown in Fig.~\ref{Fig2}.
The thermal current autocorrelation function is strictly time-independent  in the chain limit $J_\perp=0$ \cite{zotos97} (data not shown in the figure) but decays to zero for any $J_\perp>0$, which is consistent with earlier studies that suggested the absence of ballistic contributions in spin-ladder systems \cite{hm03,zotos04}. For the isotropic ladder $\Delta=J_\perp/J=1$ at high temperatures, this decay takes place on a fairly short time scale $t J\lesssim 8$ [see Fig.~\ref{Fig2}(a)]. In Figs.~\ref{Fig2}(b,c), we compare the behavior of $C_{\rm s}(t)$ on chains to isotropic ladders ($J_\perp=J$) for two different exchange anisotropies $\Delta=0.5$ and $\Delta=1$. In both cases, $C_{\rm s}(t)$ decays much faster if $J_\perp>0$, and our data suggest the absence of ballistic contributions to spin transport in agreement with Refs.~\onlinecite{hm03,steinigeweg14a}. Moreover, oscillations in $C_{\rm s}(t)$ emerge in the case of ladders. They become very pronounced at lower temperatures and are related to the existence of a spin gap for $J_\perp>0$.

\subsection{Extraction of conductivities} 
\label{sec:extract}

\begin{figure}[t]
\includegraphics[width=0.90\columnwidth]{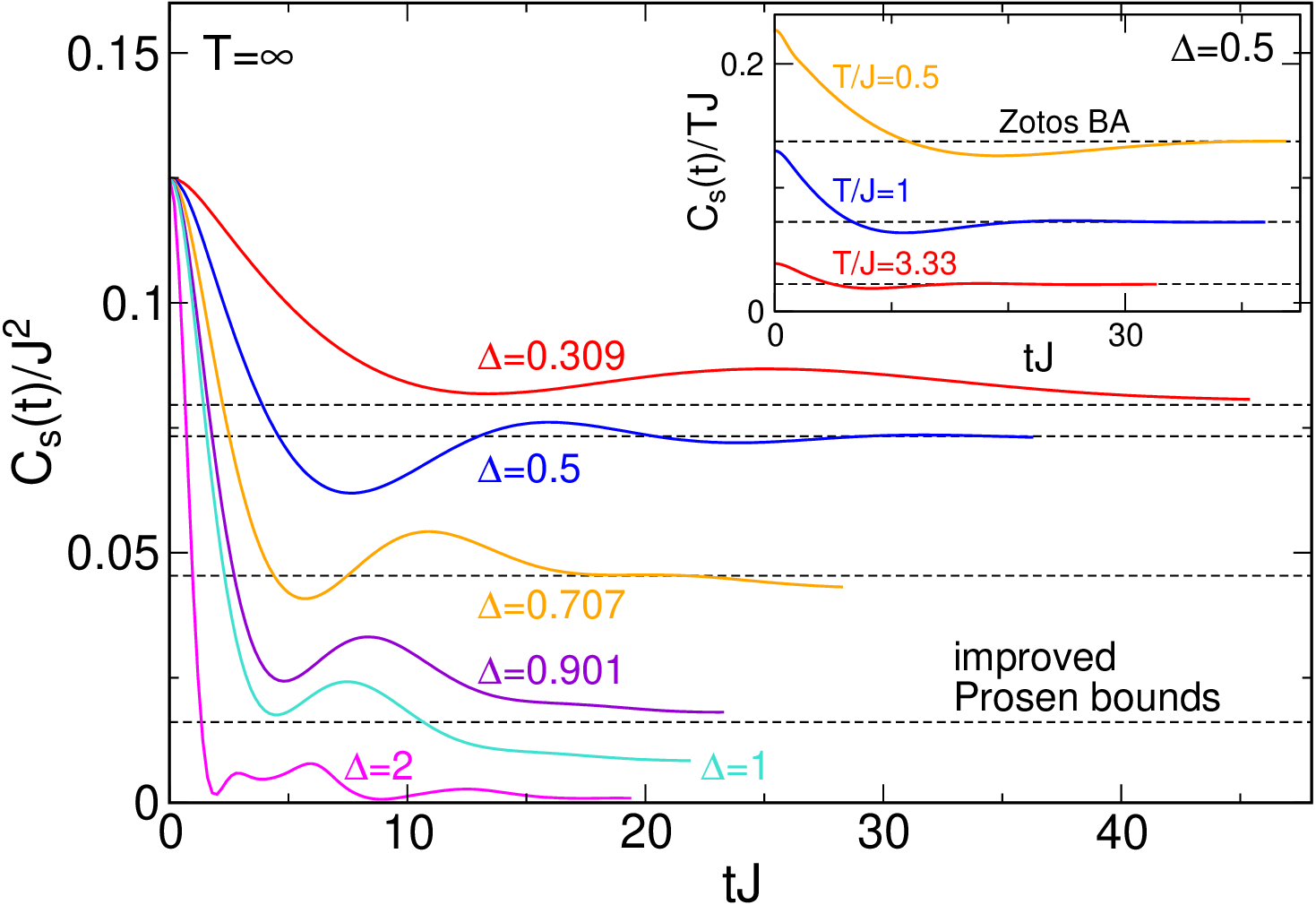}
\caption{(Color online) Real-time spin current correlation functions of the XXZ chain [see Eq.~(\ref{eq:hxxz})] at infinite temperature $T=\infty$ (main panel) and fixed exchange anisotropy $\Delta=0.5$ (inset). The model is integrable, and the spin Drude weight $D_\tn{s}$ is finite for $|\Delta|<1$ \cite{herbrych11,prosen11,karrasch13,prosen13}. The horizontal lines show the lower bounds for $D_\tn{s}$ established in Ref.~\onlinecite{prosen13} as well as the Bethe-ansatz result from Ref.~\onlinecite{zotos99}.
}\label{Fig1}
\end{figure}

We compute the spin and heat conductivities from the corresponding real-time current correlation functions via Eq.~(\ref{eq:sigma1}). However, only finite times $t<t_{\rm max}$ can be reached in the DMRG calculation of $C_\tn{s,th}(t)$, which gives rise to a `finite-time error' in $\sigma_\tn{reg}(\omega)$ and $\kappa_\tn{reg}(\omega)$. We assess this error as follows.

Our data suggest (in agreement with the results of Refs.~\onlinecite{hm03,zotos04,steinigeweg14a}) that for any $J_\perp>0$ the spin and thermal Drude weights vanish; the current correlators decay to zero for $t\to\infty$. We first compute the frequency integral in Eq.~(\ref{eq:sigma1}) using only the finite-time data. Thereafter, we extrapolate $C_\tn{s,th}(t)$ to $t=\infty$ using linear prediction \cite{barthel09} and re-compute the frequency integral. Linear prediction attempts to obtain data for correlation functions of interest at times $t>t_{\rm max}$ as a linear combination of the available data for a discrete set of times points $t_n<t_{\rm max}$ (see \cite{barthel09} for details). We perform the linear prediction for a variety of different fitting parameters (such as the fitting interval) and then define the error bar as twice the largest deviation to the conductivity computed without any extrapolation at all.

For the XXZ chain with $|\Delta|>1$, the Drude weight also vanishes, and the finite-time error of $\sigma_\tn{reg}(\omega)$ can be assessed analogously to ladders. 
The same holds at $|\Delta|<1$ and $T=\infty$ where a lower bound for $D_\tn{s}$ is known analytically from Prosen's work (see the discussion in Sec.~\ref{sec:current}).
For the values of $\Delta$ considered here, this bound agrees with the Drude weight computed using other methods \cite{zotos99,herbrych11,karrasch13}; hence, we assume that it is 
exhaustive, which allows us to subtract $D_\tn{s}$ in Eq.~(\ref{eq:sigma}).

At $|\Delta|<1$ and $T<\infty$, the Drude weight needs to be extracted from the numerical data \cite{karrasch12,karrasch13}, which is an additional source of error, or it has to be taken from other methods such as the Bethe-ansatz calculation of Ref.~\onlinecite{zotos99}. We estimate the corresponding uncertainty of the conductivity as follows. For $\Delta=0.5$ and $T=\infty$, $C_\tn{s}(t)$ oscillates around the value for $TD_\tn{s}$ known from the improved lower bound from \cite{prosen13} (see Fig.~\ref{Fig1}); for finite temperatures $T\in\{3.3J,J,J/2\}$, $C_\tn{s}(t)$ oscillates around the Bethe-ansatz result $D_\tn{s}^\tn{BA}$ of Ref.~\onlinecite{zotos99}. An upper as well as a lower bound $D_\tn{s}^\tn{u,l}$ can be determined from the magnitude of the oscillations. For each  $D_\tn{s}^\tn{BA}$, $D_\tn{s}^\tn{u}$, and $D_\tn{s}^\tn{l}$, we carry out the procedure used at $T=\infty$ and define the uncertainty in  $\sigma_\tn{reg}(\omega)$ as either twice the difference between the curves computed with and without extrapolation or twice the maximum difference between the curves at the different $D_\tn{s}^\tn{BA,u,l}$, whatever is larger.

\begin{figure}[t]
\includegraphics[width=0.90\columnwidth]{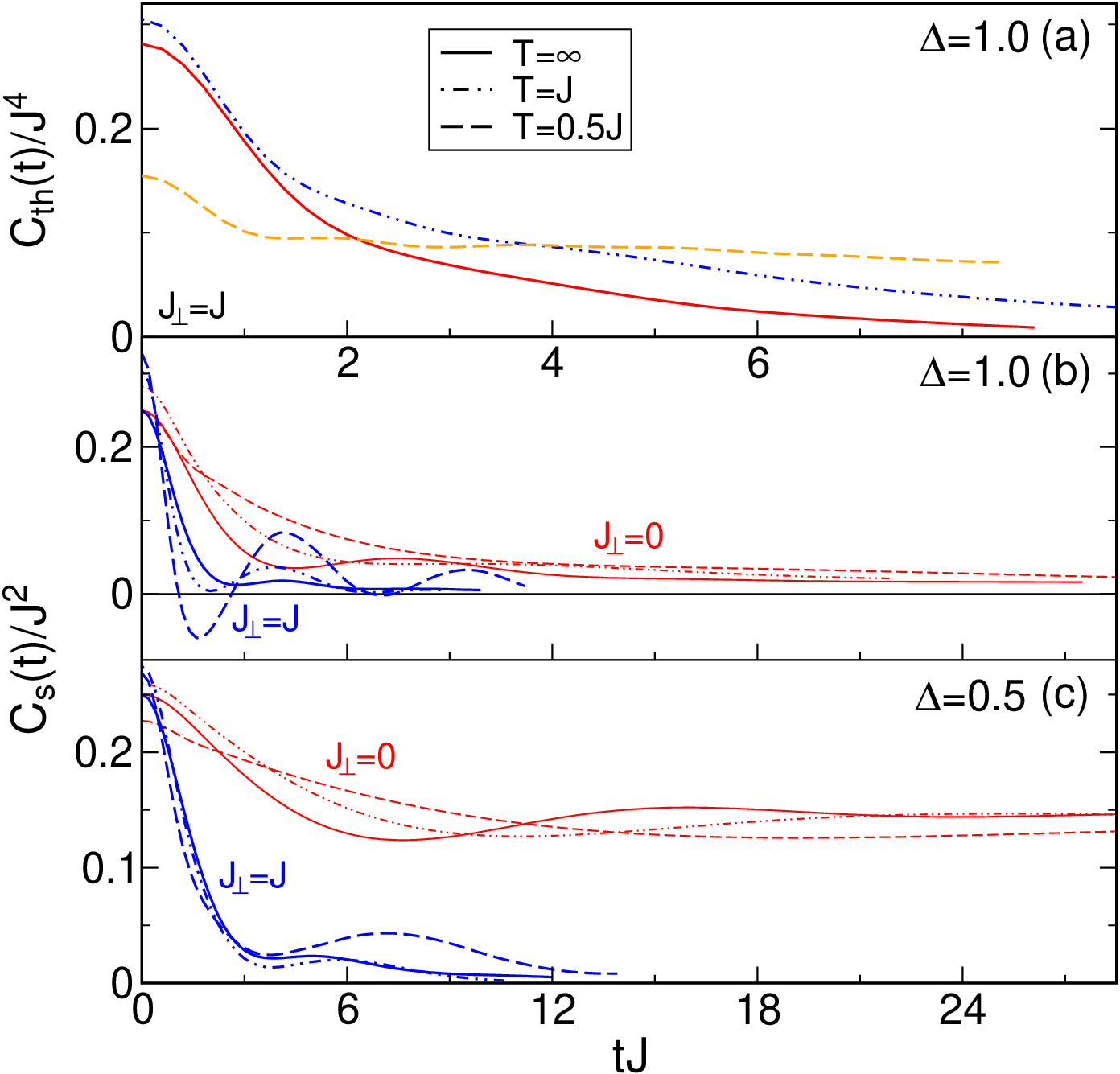}
\caption{(Color online) Current correlation functions of two-leg spin ladders governed by the Hamiltonian of Eq.~(\ref{eq:hlad}). $\Delta$ and $J_\perp$ denote the exchange anisotropy and the rung coupling, respectively. (a) Energy current  autocorrelation function of isotropic ladders $J_\perp/J=1, \Delta=1$ for various $T$. (b,c) Spin current autocorrelation functions at $\Delta=1$ and $\Delta=0.5$, respectively. For $J_\perp=0$, one recovers two identical, decoupled XXZ chains.
}\label{Fig2}
\end{figure}

For other exchange anisotropies, the accessible time scales are either too short to fully resolve the oscillations around $D_{\rm s}$, or $C_\tn{s}(t)$ decays monotonicly for large times. The latter seems to be true, in particular, close to the isotropic point $\Delta=1$. For $\Delta=0.901$ (see  Fig.~\ref{Fig1}) and at $T=\infty$, the value of $C_\tn{s}(t)$ at the largest time reached is approximately $10$ percent larger than the improved bound from \cite{prosen13}. For finite but not too small $T/J$, we assume that $C_\tn{s}(t)/(2T)=rD_\tn{s}$ at the maximal time reached, where we typically choose $r\sim1.2$. Given this estimate for $D_\tn{s}$, we assess the error analogously to the infinite-temperature case. Note that the larger $r$, the larger the error bars. We stress that this way of estimating the error is less controlled than in those cases for which the value of the Drude weight is known.

\begin{figure}[t]
\includegraphics[width=0.90\columnwidth]{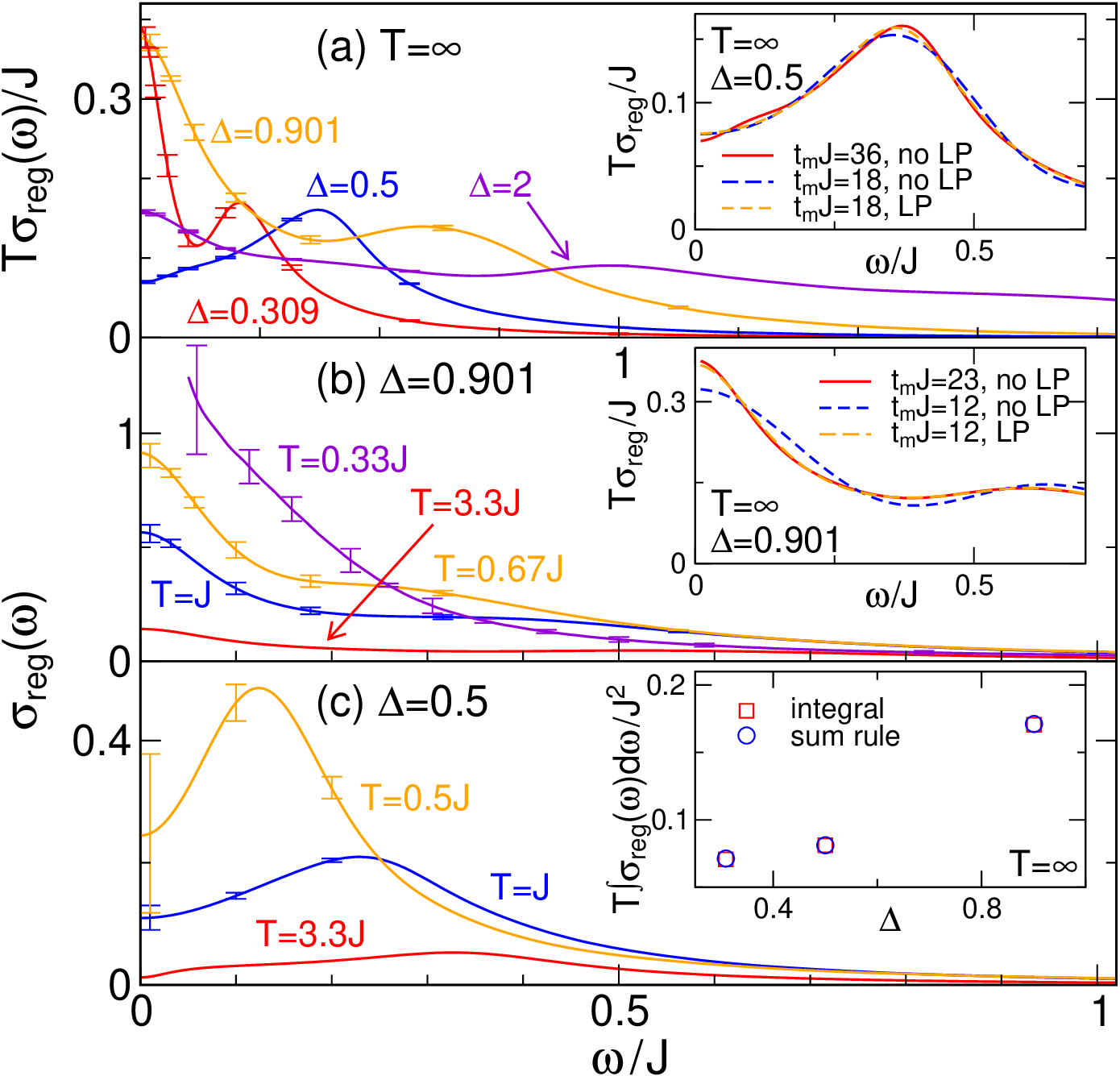}
\caption{(Color online) Regular part of the spin conductivity of the XXZ chain at (a) infinite temperature and various values of $\Delta$, and (b,c) at a fixed value of $\Delta<1$ and various $T$. The `finite-time' error can be estimated following the procedure outlined in Sec.~\ref{sec:extract}. The insets to (a) and (b) show the conductivity obtained from the finite-time data $t<t_m$ without extrapolation (`no LP') as well as from using linear prediction to extrapolate to $t\to\infty$ (`LP'). The inset to (c) illustrates that the optical sum rule Eq.~\eqref{eq:sumrule}  is fulfilled accurately.
}\label{Fig3}
\end{figure}

\begin{figure}[t]
\includegraphics[width=0.90\columnwidth]{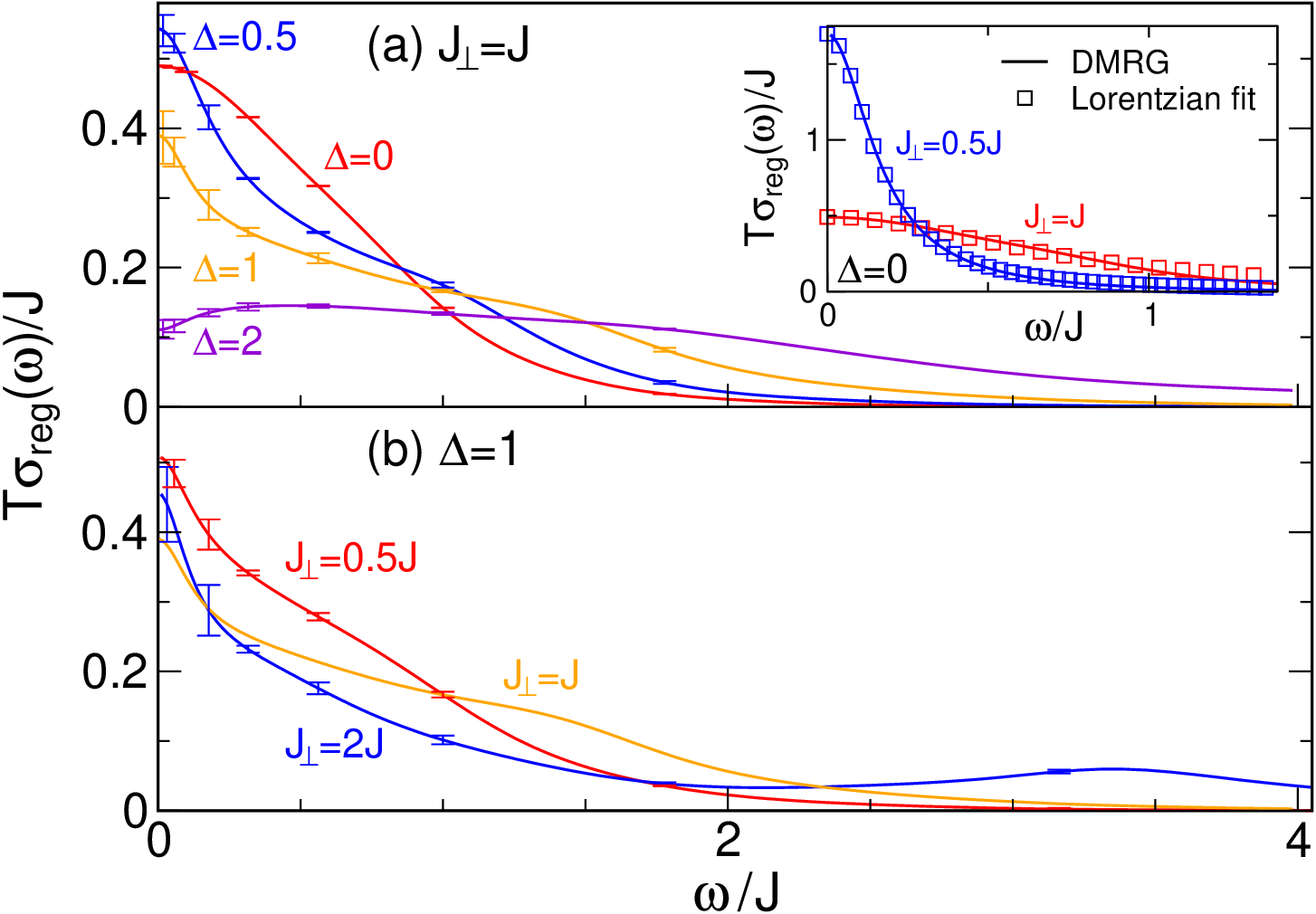}
\caption{(Color online) Spin conductivity of two-leg ladders at infinite temperature $T=\infty$ for (a) fixed rung coupling $J_\perp=J$ and (b) fixed anisotropy $\Delta=1$. At $\Delta=0$ and $J_\perp\lesssim J$, $\sigma_\tn{reg}(\omega)$ is of the simple Lorentzian form (see the inset). 
}\label{Fig4}
\end{figure}

\begin{figure}[t]
\includegraphics[width=0.90\columnwidth]{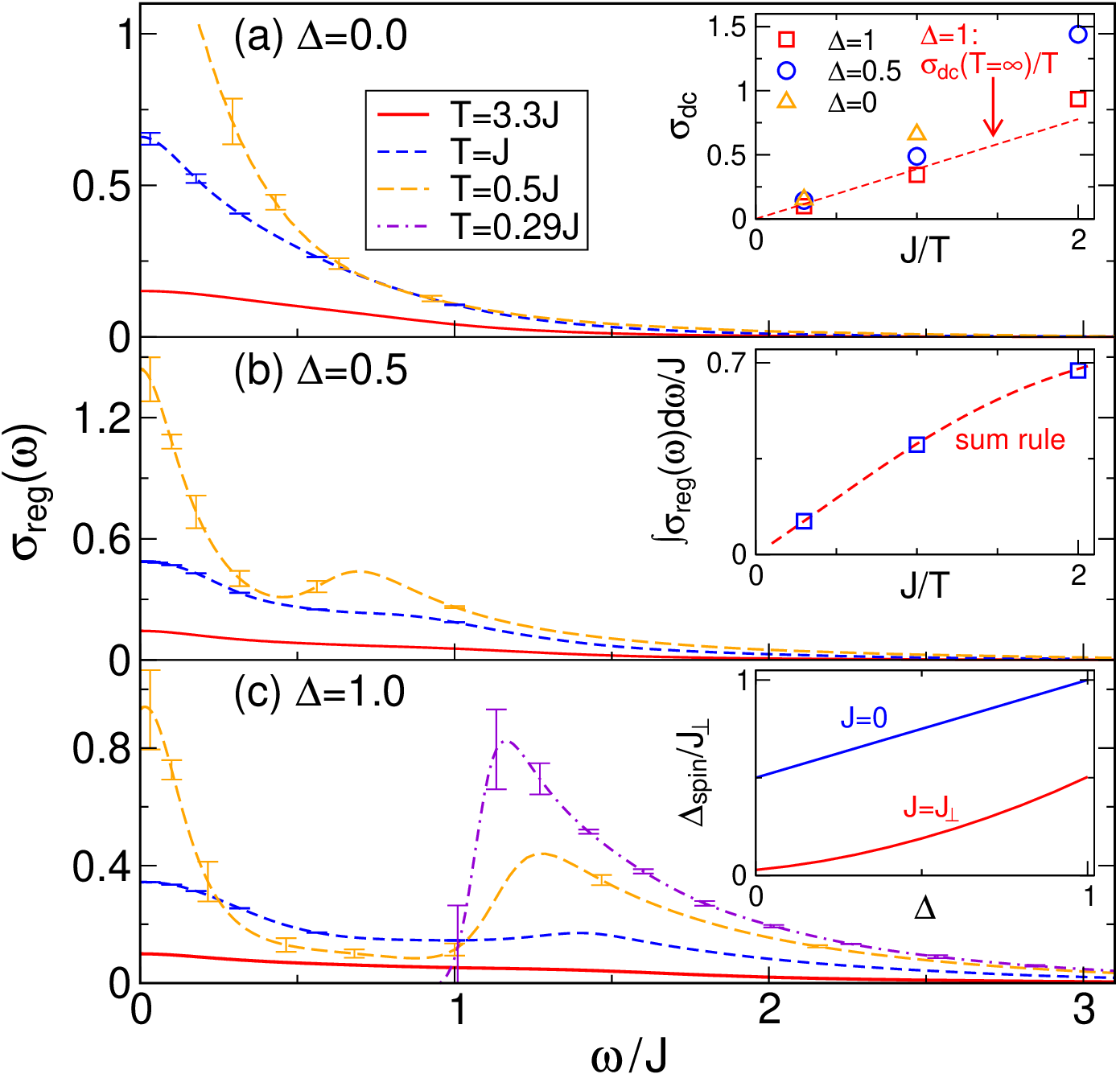}
\caption{(Color online) Spin conductivity of two-leg ladders with fixed $J_\perp=J$ but various $T$ and anisotropies ranging from $\Delta=0$ (XX ladder) to $\Delta=1$ (isotropic ladder). Note that the curve at $\Delta=1$, $T=0.29J$ (at $\Delta=0$, $T=0.5J$) is plotted only for frequencies $\omega\geq J$ ($\omega\geq 0.2J$). The insets show the DC conductivity, the optical sum rule, and the $\Delta$-dependence of the spin gap (calculated for $L=128$ at $J>0$), respectively.}
\label{Fig5}
\end{figure}

Exemplary error bars are shown in Figs.~\ref{Fig3}-\ref{Fig6}. The data for $\sigma_\tn{reg}(\omega)$ displayed in the figures are the ones obtained using linear prediction; the conductivities for the XXZ chain at $\Delta=0.5$ and $T<\infty$ shown in Fig.~\ref{Fig3}(c) were calculated using the Bethe-ansatz value of Ref.~\onlinecite{zotos99} for the Drude weight. Note that the numerical error of the bare DMRG data for $C_\tn{s,th}(t)$ is negligible compared to the finite-time error.

The finite-time data used for the above procedure is the DMRG data up to the maximum time $t_{\rm max}$ reached in the simulation. In the case of the XXZ chain, this time is fairly large compared to $1/J$, and it is thus instructive to re-calculate $\sigma_\tn{reg}(\omega)$ using only the data for half of the maximum time (both with and without extrapolation). Results are shown in the insets to Fig.~\ref{Fig3}; they illustrate that linear prediction provides a fairly reliable way to estimate the error.

As an additional test for the accuracy of the conductivities, one can verify the optical sum rule. In the spin case it reads
\begin{equation}\label{eq:sumrule}
\int_0^\infty d\omega\, \mbox{Re}\,\sigma(\omega) = \frac{\pi\langle -\hat T \rangle }{2L}\,,
\end{equation}
where $\hat T$ is the kinetic energy, i.e., all terms in Eq.~(\ref{eq:hlad}) computed at $\Delta=0$. We show exemplary data for the validity of the sum rule in the insets of Figs.~\ref{Fig3}(c) and
\ref{Fig5}(b), which illustrate that Eq.~(\ref{eq:sumrule}) holds with great accuracy.

\subsection{Spin conductivity of the spin-1/2 XXZ chain}
Figure~\ref{Fig3} gives an overview over the behavior of $\sigma_\tn{reg}(\omega)$ for the spin-1/2 XXZ chain for  various values of the exchange anisotropy  $\Delta$ (results for $\Delta>1$ have previously been shown in Ref.~\onlinecite{karrasch14a}). At infinite temperature [see Fig.~\ref{Fig3}(a)] and for all $\Delta>0$ considered, we find a finite dc conductivity $\sigma_{\rm dc} =\lim_{\omega\to 0} \sigma_{\rm reg}(\omega)$ within the error bars of our extrapolation method. 
This is at odds with the predictions of Ref.~\onlinecite{herbrych12}, where $\sigma_{\rm reg}(\omega)\propto \omega^2$ was suggested in the low-frequency limit. For the special value of $\Delta=0.5$, however, there clearly is a suppression of 
weight around $\omega=0$ accompanied by a pronounced maximum at $\omega \approx 0.25J$. For other 
values of $\Delta<1$, $\sigma_{\rm reg}(\omega)$ seems to exhibit a global maximum at $\omega=0$ as well as
additional lower maxima at higher frequencies that shift to larger values of $\omega$ as $\Delta$ increases. The spin conductivity at $\Delta=2$ has also been 
analyzed in Ref.~\onlinecite{prelovsek04}, and large anomalous, $L$-dependent fluctuations in Re$\,\sigma(\omega)$ have been observed at low frequencies. Those are not present
in our data. 
 
Returning to the regime of $\Delta<1$, we cannot rule out that the disagreement between our result for the low-frequency behavior of the conductivity and the prediction of Ref.~\onlinecite{herbrych12} is attributed to finite-time effects. However, there is no obvious indication for this in our data: At $T=\infty$, the Drude weight is known from \cite{prosen13}, and $\lim_{T\to \infty}[T\sigma_\tn{dc}]$ is simply given by the integral of $C_\tn{s}(t)$ with $2TD_\tn{s}$ subtracted. The real-time data are shown in Fig.~\ref{Fig1}; the errors due to the finite system size and the finite discarded weight are negligible. As illustrated in the insets to Fig.~\ref{Fig3}(a,b), our extrapolation scheme using linear prediction provides a stable and meaningful way to establish finite-time error bars. As an additional test, it is instructive to assume that 
the improved lower bound -- which at $T=\infty$ and for the exchange anisotropies considered here coincides with the Bethe-ansatz result of Ref.~\onlinecite{zotos99} -- is not fully saturated. At $\Delta=0.5$, an \textit{upper} bound $D_\tn{s}^\tn{u}$ to the Drude weight can be estimated from the magnitude of the oscillations of $C_\tn{s}(t)$. Using $D_\tn{s}^\tn{u}$ instead of the  lower bound from \cite{prosen13} decreases $\sigma_\tn{dc}$ by 15\% but does not yield  $\sigma_\tn{dc}=0$.

To summarize, a vanishing dc conductivity suggested by Ref.~\onlinecite{herbrych12} could only be caused by oscillations at large times around the asymptote $2TD_\tn{s}$, which would need to cancel out the large positive contribution from times $tJ\lesssim40$. Put differently, if $\sigma_\tn{reg}(\omega)\sim\omega^2$ holds, it only holds for very small frequencies $\omega\ll J$. This is further corroborated by the fact that the optical sum rule of Eq.~(\ref{eq:sumrule}) is fulfilled accurately [see the inset to Fig.~\ref{Fig3}(c)].

Even though $\sigma_{\rm dc}>0$ is supported by our tDMRG calculation in combination with the results for $D_{\rm s}(T)$ from Refs.~\onlinecite{prosen13,zotos99},
the emergence of very narrow peaks in the data for $\Delta=0.309, 0.901$ at low frequencies should be taken with some caution. For these parameters, the time scale $tJ\lesssim t_{\rm max}=40/J$ reached in the simulation is too short to resolve potential oscillations around the long-time asymptote. A very conservative estimate of the accessible frequencies is $\omega_{\rm min}= 2\pi/t_{\rm max} \sim 0.15 J$. It is possible that redistributions of weight below $\omega_{\rm min}$ would occur if longer times were available.

The temperature dependence of $\sigma_{\rm reg}(\omega)$ is shown in Figs.~\ref{Fig3}(b,c) for two different exchange anisotropies. At $\Delta=0.901$, the global maximum is always at $\omega=0$, $\sigma_\tn{dc}$ increases with decreasing temperature, and the $\omega$-dependence seems to become smoother the smaller $T$ is. For $\Delta=0.5$, the suppression of weight at low frequencies survives down to temperatures of $T\gtrsim0.5J$ (at $T=0.5J$, the error bars become too large to draw any conclusions).

To guide our ensuing discussion of ladders, 
we summarize the $\Delta$-dependence of $\sigma(\omega)$  in the chain limit $J_\perp=0$.
At $\Delta=0$, Re$\,\sigma(\omega)= 2\pi D_s(T) \delta(\omega)$, and the perturbation $J_\perp>0$ thus
breaks both the integrability of the model and the conservation of the spin current. For $\Delta>0$, the spin current is no longer conserved even for $J_\perp=0$, which gives rise to a non-zero regular contribution $\sigma_{\rm reg}(\omega)$ to the conductivity. According to recent studies \cite{prosen11,prosen13,karrasch13,herbrych11}, the Drude weight is finite for any $0\leq|\Delta|<1$, but no final conclusion on $D_\tn{s}(T)$  at  $\Delta=1$ has been reached yet. At $T=\infty$, the relative contribution of $\sigma_{\rm reg} (\omega)$ to the total spectral weight increases monotonicly from zero at $\Delta=0$ to a value of the order of 90\% close to $\Delta=1$ \cite{karrasch13}. For $\Delta>1$, the commonly expected picture is that $D_\tn{s}(T>0)=0$; hence, all weight is concentrated in the regular part. Based on these qualitative differences of $\sigma(\omega)$ that depend on $\Delta$ and the interplay of the ballistic contribution with finite-frequency weight at small $\omega$, we expect significant changes in the spin conductivity of ladders as a function of $\Delta$.

\subsection{Spin conductivity of ladders}
We now turn to the spin conductivity $\sigma(\omega)$ of two-leg ladders and contrast our results to the limiting case of isolated chains ($J_\perp=0$), where the behavior of $\sigma(\omega)$ crucially depends on $\Delta$. We first discuss the infinite-temperature case; data for $J_{\perp}=J$ are presented in Fig.~\ref{Fig4}(a). At $\Delta=0$, $\sigma_{\rm reg}(\omega)$ has a simple Lorentzian shape [see the inset to Fig.~\ref{Fig4}(a)]:
\begin{equation}\label{eq:lorentz}
\mbox{Re}\, \sigma(\omega) = \frac{\pi\sigma_{\rm dc}/\tau^2}{\omega^2 + (1/\tau)^2}\,.
\end{equation}
This follows directly from the results of Ref.~\onlinecite{steinigeweg14a}, where the spin-autocorrelation function of the XX two-leg ladder was studied numerically  and analytically as a function of $J_{\perp}/J$. It turned out that $C_\tn{s}(t)$ decays exponentially at small values of $J_\perp\lesssim J$ and with a Gaussian for larger values of $J_\perp$. The results of Ref.~\onlinecite{steinigeweg14a} in conjunction with our data altogether identify the XX spin-1/2 ladder as a textbook realization of a diffusive conductor with a {\it single} relaxation time $\tau\propto (J/J_\perp)^2$. Systems with  $\Delta=0$ are rarely found in real materials, but the XX model on a ladder can easily be realized with hard-core bosons in optical lattices (see, e.g., Ref.~\onlinecite{ronzheimer13} and the discussion in Refs.~\onlinecite{steinigeweg14a,vidmar13}).

For the special case of $\Delta=1$, we show exemplary data for $J_{\perp} \not= J$ in Fig.~\ref{Fig4}(b).
Even at $T=\infty$, the conductivity does not have a simple functional form but features  side maxima at finite frequencies that shift to larger $\omega$ as $J_{\perp}/J$ increases.

In Figs.~\ref{Fig5}(a)-(c), we illustrate how $\sigma_{\rm reg}(\omega)$ of isotropic ladders $J_{\perp}=J$ evolves as the temperature decreases from $T=\infty$ down to $T=0.29J$. It turns out that it is easier to reach low temperatures for larger values of $\Delta$. In the case of $\Delta=0$ [see Fig.~\ref{Fig5}(a)], we observe a Drude-like conductivity 
down to temperatures of $T\sim 3J$. At lower temperatures, however, Re\,$\sigma(\omega)$ deviates from a simple Lorentzian (see Ref.~\onlinecite{steinigeweg14b} for similar observations for a chain with a staggered field). This is a  consequence of the existence of a spin gap $\Delta_{\rm spin}$ in the two-leg ladder which at low
temperatures manifests itself by a suppression of weight below the optical $2\Delta_{\rm spin}$ (see, e.g., the case of dimerized chains studied in Ref.~\onlinecite{langer10}) and a sharp increase of $\sigma_{\rm reg}(\omega)$ at  $\omega \sim 2\Delta_{\rm spin}$. As a consequence, the dc conductivity is expected to diverge with $T^{-\alpha}$, $\alpha>0$ as $T$ is lowered \cite{sachdev97,damle98,damle05,karrasch14a}. Next, we investigate how the Drude-like conductivity observed for $\Delta=0$ evolves as $\Delta$ increases.
We find that (i) the current autocorrelations at $\Delta=0.5$ and $\Delta=1$ do not follow a simple exponential or Gaussian decay even at infinite temperature, and hence (ii) the low-frequency conductivity is not well-described by a simple Lorentzian. Pragmatically,  we associate the  (zero-frequency) current relaxation time $\tau$ with the inverse of the half-width-half-maximum of the zero-frequency peak in Re$\,\sigma(\omega)$ for $\Delta \gg 0$. 

The presence of these two scales, the optical gap $2\Delta_{\rm spin}$ and the inverse high-temperature relaxation time $1/\tau$, which controls the low-frequency behavior, is more visible
in the data for $\Delta=0.5$ and $\Delta=1$ [shown in Figs.~\ref{Fig5}(b) and (c)] even at the highest temperatures $T=3.3J$. 
Clearly, there are two maxima in Re$\,\sigma(\omega)$, one at $\omega=0$ and one at $\omega \gtrsim 2\Delta_{\rm spin}$ (in fact, at very low $T$, Re$\,\sigma(\omega)$ has a an edge at 
the optical gap).
The reason is the dependence of the optical gap on the exchange anisotropy $\Delta$.
The spin gap in a two-leg ladder as a function of $\Delta$ is, in the limit of $J=0$, given by  
\begin{equation}
\Delta_{\rm spin} =  \frac{J_\perp} {2} (1+\Delta)\,.
\end{equation}
This monotonic dependence of $\Delta_{\rm spin}$ on $\Delta$ survives at finite values of $J_\perp\sim J$. 
This is shown in the inset of Fig.~\ref{Fig5}(c), 
which has been obtained from $\Delta_{\rm spin} = E_0(S^z=1)-E_0(S^z=0)$ 
using standard  DMRG \cite{white92, schollwoeck05}, where $E_0(S^z)$ is the ground state in the 
subspace with total magnetization $S^z$ for $L=128$. 

\begin{figure}[t]
\includegraphics[width=0.90\columnwidth]{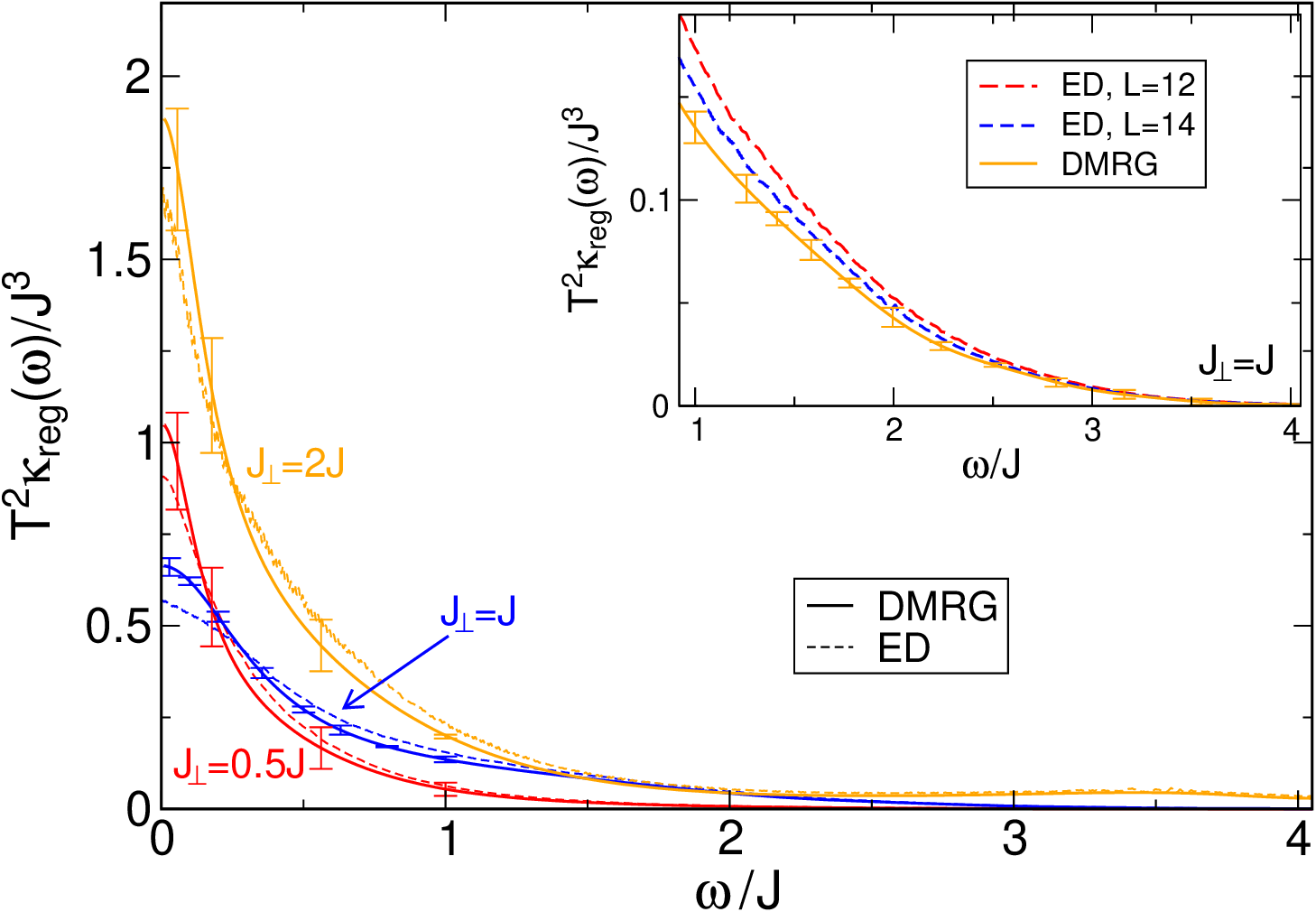}
\caption{(Color online)
Thermal conductivity of two-leg ladders at fixed $\Delta=1$ and $T=\infty$ but various $J_\perp$. We compare our data with the exact diagonalization result of Ref.~\onlinecite{zotos04}.
}\label{Fig6}
\end{figure}

Our data are compatible with a leading temperature dependence of the form $\sigma_{\rm dc}(T) \propto 1/T$.
Moreover, $\sigma_{\rm dc}$ is a monotonicly decreasing function of $\Delta$ in the high-temperature regime.
The latter can be understood by the nature of the  single-particle spin-1 excitations  of the two-leg ladder that originate from the
local triplet excitations of the $J/J_\perp\to 0$ limit. Finite values of $J$ render these triplets dispersive and give rise to interactions between the quasi-particles.
A nonzero value of $\Delta$ introduces additional scattering terms, and it is thus intuitive to expect smaller quasi-particle life times and hence also smaller dc-conductivities. 

\subsection{Thermal conductivity of Heisenberg ladders}
For experiments with quantum magnets, the thermal conductivity is the most easily accessible
transport coefficient, which has been investigated in a large number of experiments on ladders \cite{sologubenko00,hess01}, chains \cite{solo00a,sologubenko07a,hess07a,hlubek10}, and two-dimensional antiferromagnets  \cite{hess03,sales02} (see Refs.~\onlinecite{hess07,sologubenko07} for a review).
These experiments have clearly established that magnetic excitations can dominantly contribute to the thermal conductivity of these
insulating materials at elevated temperatures,  exceeding the phononic contribution (see, e.g., \cite{hlubek10}).
The contribution of magnetic excitations to the full thermal conductivity in these low-dimensional systems manifests itself via
a prominent anisotropy of the thermal conductivity measured along different crystal axes \cite{hess07,sologubenko07}.
Open and timely questions include a comprehensive and quantitative theoretical explanation for the magnitude of the thermal conductivity,
a theory of relevant scattering channels beyond pure spin systems (see, e.g., \cite{shimshoni03,chernyshev05,rozhkov05a,boulat07,gangadharaiah10,bartsch13,rezania13}), 
a full understanding of the 
spin-phonon coupling including spin-drag effects \cite{boulat07,bartsch13,gangadharaiah10}, and the understanding of a series of experiments studying the effect of  doping
with nonmagnetic or magnetic impurities 
and disorder onto the thermal conductivity (see, e.g., \cite{hess04,hess06}).
Here we solely focus on pure spin Hamiltonians.
Given that most of the materials realize spin Hamiltonians that are more complicated than just spin chains with nearest-neighbor interactions 
only, one needs to resort to numerical methods to get a quantitative picture.

The real-time energy current autocorrelations for the Heisenberg ladder ($\Delta=1$) are shown in Fig.~\ref{Fig2}(a). At $T=\infty$, $C_{\rm th}(t)$ decays fast, and $\kappa(\omega)$ can be obtained down to sufficiently low frequencies. For lower temperatures, however, the accessible time scales are at present too short to reach the dc-limit in a reliable way. We therefore focus on $T=\infty$.

Our results for the thermal conductivity are shown in Fig.~\ref{Fig6} for $J_\perp/J=0.5,1,2$. They are in  reasonable agreement with the exact diagonalization data of Ref.~\onlinecite{zotos04} that were obtained using a micro-canonical Lanczos method for $L=14$ sites. Note that our data for $\kappa_{\rm dc}$ is typically larger than the ED results. The behavior at low frequencies is anomalous -- 
it does not follow a Drude-like Lorentzian shape (this was already pointed out in Ref.~\onlinecite{zotos04}). The actual form of the 
low-frequency dependence of  $\kappa_{\rm reg}(\omega)$  (discussed in Ref.~\cite{zotos04}) cannot be clarified using the existing data.
The knowledge of the infinite temperature dc-conductivity $\kappa_{\rm dc}^{\infty}$ still gives access to a wide temperature regime since the leading term is $\kappa_{\rm dc}(T) = \kappa_{\rm dc}^{\infty}/T^2$, and we can therefore address the question of mean-free paths.

\subsection{Mean-free paths for the Heisenberg spin ladder}
In the analysis of experimental data for $\kappa$, one often uses a kinetic equation to extract magnetic mean-free paths \cite{hess01,hess07}.
An analogous equation can also be employed for $\sigma$, and we obtain the following set of kinetic equations:
\begin{eqnarray}
\kappa &=&  \frac{1}{L}\sum_k v_k \frac{d(\epsilon_k n_k)}{dT} l_{\kappa,k} \label{eq:lkappa}\\
\sigma &=& \frac{1}{L}\sum_k v_k  \left(-\frac{dn_k}{d\epsilon_k}\right) l_{\sigma,k}\,,  \label{eq:lsigma}
\end{eqnarray}
where $\epsilon_k$ is the dispersion of the threefold degenerate triplet excitations, $v_k =\partial_k \epsilon_k$,
and $n_k$ is a distribution function which accounts for the hard-core boson nature of the triplets \cite{hess01}:
\begin{equation}
n_k = \frac{3}{\mbox{exp}(\beta \epsilon_k)+3}\,.
\end{equation}
The actual form of the dispersion is not important since $v_k$ drops out in one dimension when the integration over $k$ is replaced by an integral over energy $\epsilon$.
The mean-free paths $l_{\kappa,k}$ and $l_{\sigma,k}$ are taken to be independent of quasi-momentum, $l_{\kappa(\sigma),k}=l_{\kappa(\sigma), \rm mag}$.
In order to analyze the total thermal conductivity measured experimentally, one assumes $\kappa_{\rm total } = \kappa_{\rm ph} + \kappa_{\rm mag}$,
where $\kappa_{\rm ph}$ and $\kappa_{\rm mag}$ represent the phononic and magnetic contribution, respectively. 
Such a separation is an approximation and should be understood as an operational means to extract mean-free paths -- in general, spin-drag effects can lead to additional contributions to $\kappa_{\rm total }$ \cite{chernyshev05,boulat07,bartsch13}.

In the high-temperature limit, one needs to keep only the leading terms in a $1/T$ expansion of Eqs.~\eqref{eq:lkappa} and \eqref{eq:lsigma}. The mean-free paths can then be extracted from $\kappa_{\rm dc} = \kappa_{\rm dc}^\infty/T^2$ and $\sigma_{\rm dc} = \sigma_{\rm dc}^\infty/T$ via  
\begin{eqnarray}
\kappa_{\rm dc}^{\infty} = \frac{1}{16\pi} (\epsilon_{\rm max}^3 -\epsilon_{\rm min }^3) l_{\kappa,\rm mag }\,, \\
\sigma_{\rm dc}^{\infty} = \frac{3}{4 \pi}  (\epsilon_{\rm max} -\epsilon_{\rm min}) l_{\sigma,\rm mag }\,,
\end{eqnarray}
where $\epsilon_{\rm max}$ and $\epsilon_{\rm min}$ are the band minimum and band maximum of the single-triplet dispersion, respectively.
For an isotropic ladder system such as the one realized in La$_5$Ca$_{9}$Cu$_{24}$O$_{41}$ (the actual Hamiltonian is more complicated, though \cite{notbohm07}), 
$\epsilon_{\rm min } =\Delta_{\rm spin}\approx J/2$ and $\epsilon_{\rm max} \approx 2J$ are reasonable estimates \cite{knetter01} for $J_\perp = J$. We can thus approximate $\epsilon_{\rm max} -\epsilon_{\rm min}=  3J/2$ and $\epsilon_{\rm max}^3 -\epsilon_{\rm min }^3 \approx 8 J^3$, which then leads to
\begin{eqnarray}
\kappa_{\rm dc}^{\infty} = \frac{J^3}{2\pi}  l_{\kappa,\rm mag }\,, \\
\sigma_{\rm dc}^{\infty} = \frac{9 J}{8 \pi}  l_{\sigma,\rm mag }\,.
\end{eqnarray}
For the isotropic ladder $J_\perp/J=1, \Delta=1$, we have $\kappa_\tn{dc}^\infty\approx0.66J^3$ and $\sigma_\tn{dc}^\infty\approx0.39J^2$ and thus $l_{\kappa,\tn{mag}}\approx 4.2 $ and $l_{\sigma,\tn{mag}}\approx 1.1$. Hence, $ l_{\kappa,\tn{mag}} > l_{\sigma,\tn{mag}}$ such that the  (averaged) mean-free paths differ from each other.
In this framework the mean-free paths in the high-temperature regime are $T$-independent, which seems
reasonable since at large $T\gg J,J_{\perp}$ (i.e., $T$ larger than the band-width of triplets)  all states are populated equally.
In other words, the qualitative difference with phonons, the most typical bosonic quasi-particle that contributes to the thermal conductivity in solids,  is that the number of triplet excitations saturates at large $T$
due to their hard-core nature, reflecting the fact that the spin system has a spectrum that is bounded from above.

Our results demonstrate that the extraction of mean-free paths as commonly employed in the analysis of the experimental data, while providing very useful
intuition, cannot easily be related to single-excitation mean-free paths, due to the different results obtained for $\kappa$ and $\sigma$
and the gap in the excitation spectrum (see also the discussion in Ref.~\onlinecite{chernyshev05}). We stress that the observation of different mean-free paths for different transport channels is not  unusual. Even in metals (more generally, 
Fermi-liquids) momentum and energy can relax differently via inelastically scattering processes \cite{ashcroft}. Moreover, more dramatic deviations from the Wiedemann-Franz law 
are well-known for non-Fermi-liquids (see, e.g., \cite{hill}),  in  Luttinger liquids \cite{kane96,wakeham11} and mesoscopic systems \cite{vavilov,kubala08}. 

\section{Summary}
\label{sec:sum}
In this work, we studied the spin and thermal conductivity of spin chains and ladders using finite-temperature, 
real-time density matrix renormalization group techniques. We first computed the spin conductivity
of the spin-1/2 XXZ chain as a function of the exchange anisotropy $\Delta>0$.  
Our data suggest finite dc-conductivities for all $\Delta>0$, yet a suppression of 
weight at low frequencies for special values such as $\Delta=0.5$. While the main drawback
of the numerical method is that only finite times can be reached in the simulations, the comparison
of various schemes to extract the frequency dependence supports our conclusion.

Our results for two-leg spin ladders are consistent with the absence of ballistic contributions in agreement with Refs.~\onlinecite{hm03,zotos04,jung06,steinigeweg14a}.
At high-temperatures, the XX ladder -- which is equivalent to a system of hard-core bosons -- exhibits a simple,
Drude-like spin conductivity \cite{steinigeweg14a}. 
This property is lost as either the temperature is lowered or the exchange anisotropy is increased.
At low temperatures, the spin conductivity features a two-peak structure with a maximum at $\omega=0$
and a large weight for frequencies above the optical spin gap.
We further computed the dc spin conductivity; it decreases as the exchange anisotropy increases from $\Delta =0$ towards $\Delta=1$
and is a monotonicly increasing function of temperature.

The thermal conductivity was obtained in the infinite-temperature limit, and our data agree reasonably well with earlier exact diagonalization results
\cite{zotos04}. We extracted estimates for mean-free paths via kinetic equations that are used in the analysis of experimental data \cite{hess01}. 
The (momentum-averaged) mean-free paths $l_\tn{mag}$ obtained from $\kappa$ are larger than the ones calculated from $\sigma$. Thus, $l_\tn{mag}$ depends on the type of transport considered, and it is therefore not obvious that values for $l_\tn{mag}$ can directly be interpreted as a mean-free path of single-particle excitations. Future time- and real-space experiments could provide additional insight into the connection between single-excitations and the mean-free paths observed in transport measurements.

{\it Acknowledgment.} We thank W. Brenig, P. Prelov\v{s}ek, T. Prosen, R. Steinigeweg, and X. Zotos for very useful discussions. We are further indepted 
to X. Zotos for sending us exact diagonalization data from Ref.~\onlinecite{zotos04} and Bethe ansatz results  for $D_{\rm s}(T)$ computed with the methods of Ref.~\onlinecite{zotos99} for comparison. We acknowledge support by the Nanostructured Thermoelectrics program of LBNL (C.K.) as well as by the DFG through the Research Training Group 1995 (D.M.K) and through FOR 912 via grant HE-5242/2-2 (F.H.-M.).

\bibliography{references}

\end{document}